\documentclass[usenatbib]{mn2e}
\usepackage{graphicx}
\usepackage[usenames]{color}
\usepackage{times}
\usepackage{rotating}
\usepackage{lastpage}
\def\gtsim{\lower.5ex\hbox{$\; \buildrel > \over \sim \;$}}
\def\ltsim{\lower.5ex\hbox{$\; \buildrel < \over \sim \;$}}

\title[Isolated GAMA galaxies]{Galaxy And Mass Assembly (GAMA): The Most Isolated GAMA Galaxies}
\author[S. Phillipps et al.]{S. Phillipps$^1$, R. De Propris$^{2,3}$, M.N. Bremer$^1$, P.A. James$^4$ and A.M. Hopkins$^5$\\
$^1$Astrophysics Group, School of Physics, University of Bristol, Tyndall Avenue, Bristol, BS8 1TL, UK\\
$^2$Finnish Centre for Astronomy with ESO, University of Turku, Finland, Vesilinnantie 5, FI-21400, Turku, Finland\\
$^3$Department of Physics and Astronomy, Botswana International University of Science and Technology, Private Bag 16, Palapye, Botswana\\
$^4$Astrophysics Research Institute, Liverpool John Moores University, IC2, Liverpool Science Park, 146 Brownlow Hill, Liverpool L3 5RF, UK\\
$^{5}$School of Mathematical and Physical Sciences, 12 Wally's Walk Macquarie University, Sydney, NSW2109, Australia\\
}

\begin{document}

\date{Accepted . Received ; in original form }

\pagerange{\pageref{firstpage}--\pageref{lastpage}} \pubyear{}

\maketitle

\label{firstpage}

\begin{abstract}
While galaxies generally live in clusters and groups, particularly small groups, or other structures such as walls or filaments, some galaxies appear relatively `isolated', i.e. lacking in close neighbours. Here we use the GAMA survey, and an objective measure of isolation, to explore the most isolated galaxies. We then determine whether these have particular physical characteristics. We find that, as expected, while grouped galaxies split between blue, disk-like and star forming and red, spheroidal and quiescent, isolated galaxies are preferentially the former. However, there is a broad overlap in properties, with some isolated galaxies sharing the latter characteristics. Considering only the star-forming galaxies, when controlled by mass we find no differences between the grouped and isolated galaxies in terms of their star formation and stellar populations.  Given the lack of significant interactions for the isolated sample, this implies that the evolution of present-day star-forming galaxies in all environments (at least outside rich clusters) has been, and remains, dominated by internal processes.

\end{abstract}


\section{Introduction}
It has long been known that galaxies congregate in clusters \citep{Wolf1902} and groups \citep{Shapley1932}, particularly small groups akin to the Local Group \citep{Holmberg1940, deV1971}. Studies of the distribution of galaxy properties, originally luminosities (i.e., the luminosity function), for galaxies in different environments also have a long history \citep{Hubble1931, Shapley1933}.

Many modern studies have considered galaxy properties separately for cluster galaxies and for non-cluster (`field') galaxies. The latter generally include (apparently) isolated galaxies \citep[e.g.,][]{Karachentseva1973} and those in small to medium sized groups \citep[e.g.,][]{TG1976}, as well as more recent categorisations such as galaxies in `filaments' or `tendrils' \citep{Alpaslan2014, Alpaslan2015}, and so-called `void galaxies', i.e. those in the lowest global density environments \citep[e.g.,][]{Rojas2004, Porter2023}. For instance, it is well known that earlier type galaxies are found in locally denser environments \citep{Hubble1931, Dressler1980}. This and many other observed differences between cluster galaxies and field galaxies are generally explained by evolutionary processes, such as star formation, galaxy interactions or mergers, occurring in different ways (or with different relative strengths) in different environments \citep[e.g.,][and references therein]{Dressler1984, Brough2013, Davies2016}. 

With the advent of huge spectroscopic data sets such as SDSS \citep[e.g.,][]{York2000, Aihara2011} it has become possible to examine in detail the environment of large numbers of individual galaxies, each characterised by numerous derived physical properties \citep[e.g.,][]{Kauffmann2003}. 
A problem at the low density end, however, is that different authors use different definitions and criteria for selecting field or isolated galaxies. 

Following Karachentseva's (1973) pioneering Catalogue of Isolated Galaxies (CIG; available via NED), which contained galaxies with no neighbours visible on Palomar plates within 20 galaxy diameters, and Turner \& Gott's (1976) list of galaxies not assigned to one of their groups, numerous other authors have constructed samples of isolated galaxies in various ways. For instance, \cite{Marquez1996} attempted to identify galaxies which had evolved free of external influences, finding 22 non-group galaxies with no companions closer than 0.5~Mpc in projected separation and/or 500~km~s$^{-1}$ in redshift in the CfA catalogue and also with no fainter close neighbours in the Palomar Sky Survey. \cite{Varela2004} extended this by placing a limit on the estimated ratio of tidal force from a neighbour to a galaxy's self-gravity, hence allowing for the mass ratio of companions as well as their distance. \cite{Verdes2005}, \cite{Verley2007a} and \cite{Argudo2013} returned to the CIG sample but used more modern multi-wavelength data and a combination of local galaxy density and tidal strengths to determine isolation for their AMIGA (Analysis of the interstellar Medium of Isolated GAlaxies) project, showing that CIG objects showed a continuous spectrum of isolation as quantified by these two parameters. There also exist several SDSS based catalogues of isolated objects, using projected separations and redshift differences, such as those of \cite{Hernandez2010} and \cite{Argudo2015} (the latter also used by \cite{Kinyumu2024} as part of a Digital Survey Isolated Galaxies (DSIG) catalogue). 

As is evident, many different criteria can be used for isolation and these may not coincide. For example, \cite{Verley2007b} present a catalogue of neighbours around `isolated' CIG galaxies, some of which were shown to be physical companions. Similarly, \cite{Torres2026} note that, with their specific definitions, 22\% of void galaxies (i.e. those in particularly low density environments outside of filaments etc.) still have companions. They therefore suggest considering simultaneously the effects of both large scale structure and small scale environment on galaxy properties. 

If we wish to consider the lowest possible density environments, where interactions might be considered negligible, we need to take care whether a galaxy is genuinely isolated or just appears so because of survey limits, for instance by missing any companions just below the survey magnitude limit. On the other hand, a luminous galaxy may be `grouped' but only with galaxies orders of magnitude smaller than itself, so may be effectively isolated. 

In the present paper, we explore the most isolated galaxies in the Galaxy And Mass Assembly (GAMA) spectroscopic survey \citep{Driver2022} of around 300,000 galaxies, making use of the associated catalogues of photometric and derived stellar population parameters \citep[e.g.,][]{Taylor2011, Wright2016} and grouping status, as derived from a friends-of-friends algorithm \citep{Robotham2011}. Similar data have previously been used, for example, to study galaxies classified as members of pairs \citep{Davies2015} or belonging to filaments and tendrils \citep{Molina2025}. 

Section 2 describes the GAMA spectroscopic and associated photometric data, the group catalogues and the selection of our specific samples. Section 3 presents our main results and Section 4 discusses these results. Section 5 gives a summary of our conclusions.

Where required for luminosities and masses we adopt the simple concordance cosmology, $H_0 = 70~$km~s$^{-1}$Mpc$^{-1}$, $\Omega_m = 0.3$, $\Omega_{\Lambda} = 0.7$ as used in the standard GAMA catalogues. Using a Planck Collaboration 2018 cosmology \citep{Aghanim2020} would make no noticeable difference to our results (around 3\% in distance for galaxies at the maximum $z$ of the samples we use), and none to our conclusions, which all rely on internal comparisons between galaxies which are isolated and those that are not.

\section{Data and Samples}

Our data come from the Galaxy And Mass Assembly (GAMA) spectroscopic survey and its allied multi-wavelength photometric data sets. GAMA is primarily a large, highly complete spectroscopic survey which was carried out with the AAOmega spectrograph on the Anglo-Australian Telescope. It obtained redshifts for around 300,000 galaxies in three equatorial fields (G09, G12 and G15) and two southern fields (G02 and G23), each covering $\simeq 60$ square degrees. Data releases are described in \cite{Driver2011}, \cite{Liske2015}, \cite{Baldry2018} and \cite{Driver2022}. 

The original optical photometry (and therefore selection) for GAMA was obtained from SDSS \citep[specifically their DR8;][]{Aihara2011} but in the latest releases \citep{Driver2016, Driver2022} this has been replaced by data from KiDS \citep[the Kilo-Degree Survey;][]{deJong2013}. 
KiDS is a wide-field imaging survey of the Southern sky in the optical broad-band filters $u, g, r, i$ carried out using the VLT Survey Telescope (VST) at the ESO Paranal Observatory. Similarly, the VISTA Kilo-degree Infrared Galaxy (VIKING) survey provides corresponding near-infrared data in the $Z, Y, J, H, K_s$ bands \citep{Edge2013, Wright2019}. The GAMA equatorial survey regions have been covered in KiDS DR3.0 \citep{deJong2017}. Our source of masses and rest-frame intrinsic stellar luminosities is the GAMA catalogue StellarMassesLambdarv24. The `Lambdar' photometry reduction is discussed in detail in \cite{Wright2016}. Using this photometry \citep[see][]{Bellstedt2020}, \cite{Driver2022} demonstrate that the GAMA spectroscopic sample in the equatorial regions is 95\% complete to a KiDS $r$-band magnitude limit of $r=19.65$. 

Stellar Population parameters are derived using MAGPHYS \citep{DaCunha2008, DaCunha2010} and see \cite{Taylor2011} for the methodology behind the spectral energy distribution (SED) fitting for rest frame magnitudes and masses.\! \footnote{Closely similar mass results are obtained with the newer ProSpect SED fitting routines, see \cite{Robotham2020}.} Data on galaxy groups and the galaxies they contain (as derived by a friends-of-friends analysis) are taken from the GAMA data management units (DMUs) G3CFoFGroupsv10 and G3CGalv10 \citep[see][]{Robotham2011}. Our base sample is then the triple match of the GAMA catalogues StellarMassesLambdarv24, MagPhysv06 and G3CGalv10, i.e. all galaxies with entries in all three of these catalogues. These number 183969. All our catalogue manipulation is carried out using the TOPCAT package \citep{MBTaylor}.

For our specific purposes we need a nearby sample so as to be able to reach faint group members (or, indeed, faint single galaxies). Specifically we impose an upper limit at $z=0.1$. A lower limit of $z=0.003$ is also set (most `galaxies' below this are produced by overlapping stellar images). We also remove any galaxies with (probably spurious) derived rest-frame, dust-corrected $r$-band stellar absolute magnitudes $M_{r,*}$ brighter than $-24$ or fainter than $-11$. This provides us with a sample of 20719 galaxies from the three catalogues, which we call MASSES3. Note that for our assumed cosmology\footnote{The absolute magnitudes are only affected at the hundredths of a magnitude level by the choice of cosmology.} the distance modulus at $z=0.1$ is 38.3 mag., so using the above apparent $r$-band limit of 19.65 we should be complete to an absolute magnitude limit of $-18.65$ at our maximum $z$. As the brightest galaxies in the sample have $M_{r,*} \simeq -23.5$ this gives a possible range of almost 5 magnitudes (see Fig. 1 below). Obviously the accessible range will be even larger at smaller $z$, in principle up to about 12 mag. at $z \simeq 0$, though in reality 10 mag., as we have no extreme luminosity galaxies in the small volume at low $z$. 

We now use the grouping data from \cite{Robotham2011} and divide our sample into isolated or grouped, i.e. members of any group with multiplicity 2 or higher. We refer to these subsets of isolated and grouped galaxies as MASSES3-IS0 (11406 galaxies) and MASSES3-GROUP (9259 galaxies). Note that although the linking lengths in the friends-of-friends algorithm are individually tuned for galaxies of different luminosity at different redshifts, via the local density and accessible part of the luminosity function \citep{Robotham2011}, it can be checked from the GAMA DMU G3CLinkv10 that for our $z$ range the maximum projected linking length for grouping galaxies together is strongly peaked around 300~kpc in projected separation$\!$ \footnote{Note that even the short side of the GAMA fields is 5 degrees long, corresponding to around 10 to 30 Mpc for almost all the redshift range considered, so incompleteness of the search for companions due to linking lengths reaching to the edge of the survey fields is small; depending on $z$, 3 to 8\% of galaxies are this close to an edge.} and 20 times greater in radial separation (equivalent to $cz \simeq 400~$km~s$^{-1}$) to allow for peculiar velocities. Thus non-grouped galaxies generally have no neighbours within these limits. Note that, overall, the grouped galaxies are dominated by small groups, there are no rich clusters in the GAMA regions. Indeed, for the whole of GAMA, the average number of galaxies per group is 3.1 \citep{Robotham2011}. Specifically in our sample MASSES3-GROUP, 1836 of the 2984 groups involved have multiplicity $N_{fof} = 2$ and 530 have $N_{fof} = 3$, so over half of our galaxies (5262) are in such groups. There are 93 groups with $N_{fof} > 20$ which contain 20\% of our galaxies.

The distribution of stellar (dust-corrected) absolute magnitudes as a function of redshift is shown for these two subsets in Fig. 1. We can notice immediately that both samples span the full range of accessible magnitudes. The isolated galaxies are more visible on the plot at redshifts where there are few groups but their redshift distribution is actually quite similar, for instance sharing the relatively low density in the $z$ range around 0.065, though the grouped galaxies do dominate in the densest regions (see Fig. 2).

\begin{figure}
\includegraphics[width=\linewidth]{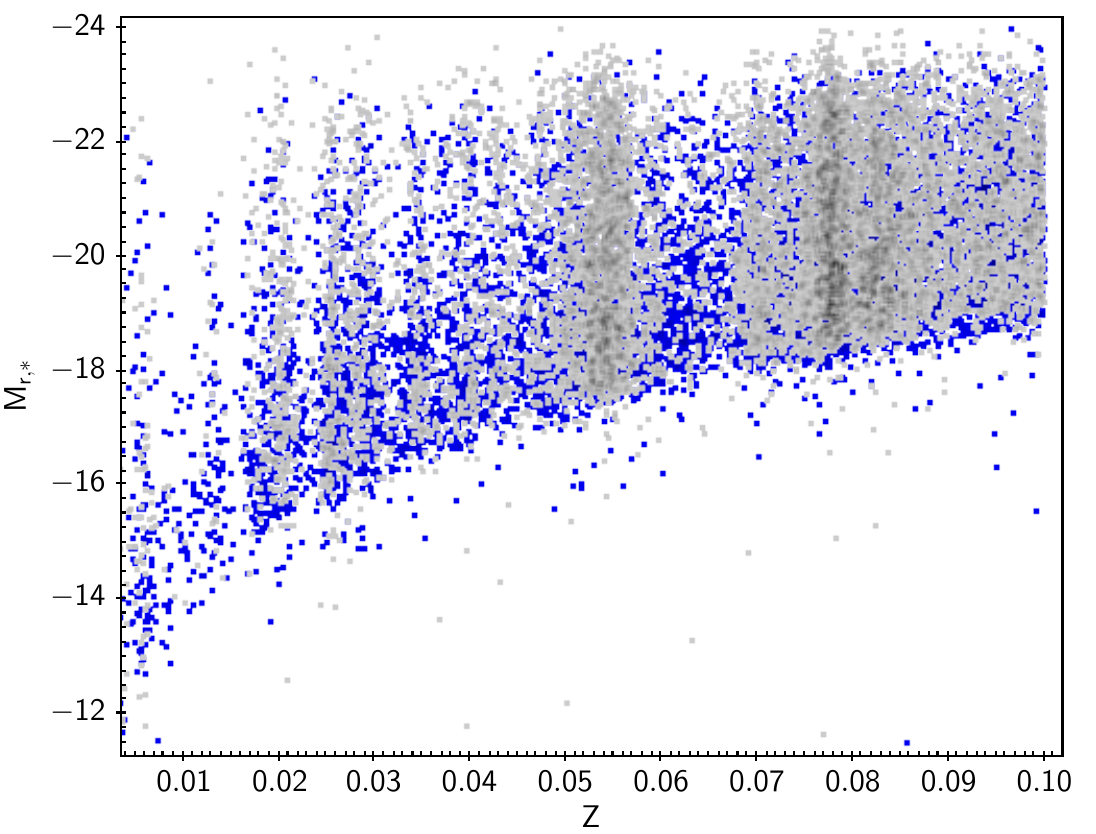}
\caption{Distribution of stellar (i.e dust-corrected) $r$-band absolute magnitude as a function of redshift for our MASSES3 sample limited to $0.003 \leq z \leq 0.1$. Grouped galaxies (MASSES3-GROUP) are shown in grey, isolated galaxies (MASSES3-ISO) in blue.
}
\label{mrz}
\end{figure}

As we wish to quantify how isolated a galaxy might be, in terms of lack of companions with which it could interact, it is easier to return to apparent magnitudes. (For consistency with Fig. 1, we calculate an effective, dust-corrected apparent magnitude, $m_{r,*}$ from $M_{r,*}$ and the distance). Taking the effective (empirical) limit to be 19.5, we can simply calculate $\Delta r = 19.5 - m_{r,*}$ (so $\Delta r$ is positive for objects above the lower bound in luminosity). The distribution of $\Delta r$ for MASSES3-ISO is shown in Fig. 3.

\begin{figure}
\includegraphics[width=\linewidth]{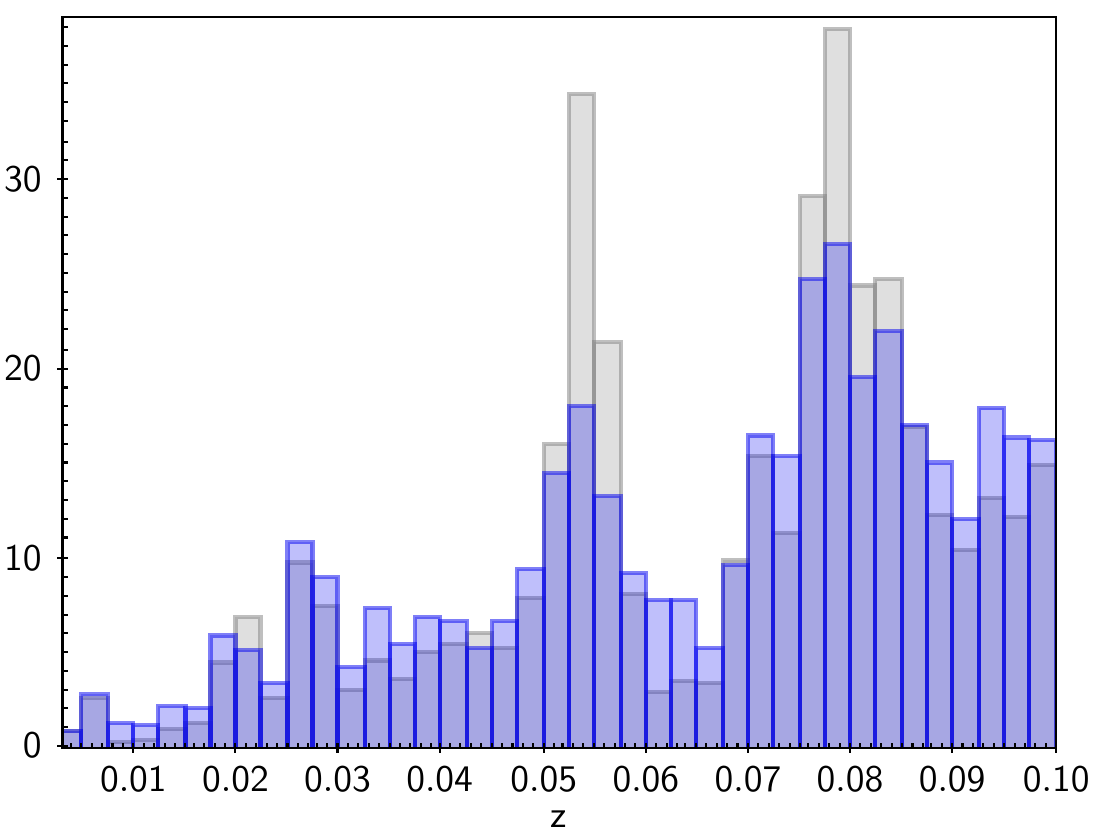}
\caption{Redshift distributions for the samples MASSES3-ISO (blue) and MASSES3-GROUP (grey), normalised by the numbers in each sample.
}
\label{iso_z}
\end{figure}

\begin{figure}
\includegraphics[width=\linewidth]{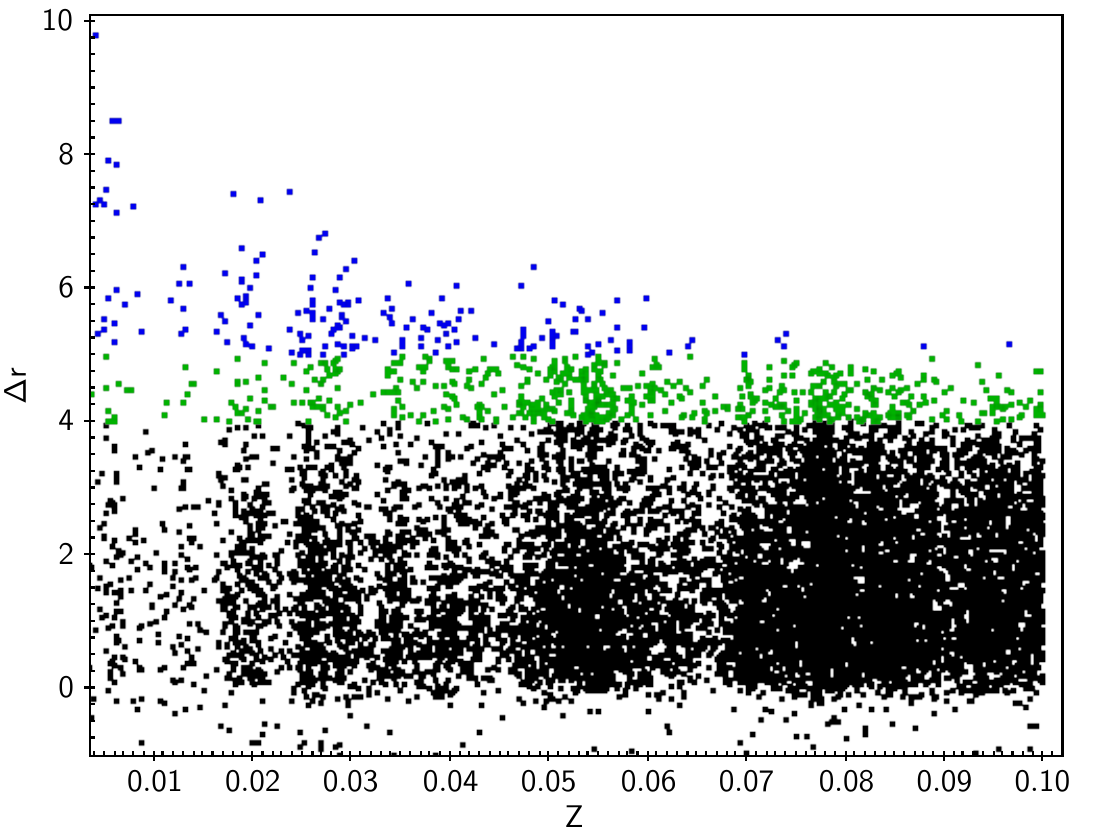}
\caption{Accessible magnitude range, $\Delta r$ for the sample MASSES3-ISO. Galaxies with $4 < \Delta r < 5$ (sample ISO4) are shown in green and those with $\Delta r > 5$ (sample ISO5) in blue. 
}
\label{iso_delta}
\end{figure}

As above, towards $z = 0.1$, there is an accessible range of about 5 mag. As our first subsample of `very' isolated galaxies, we therefore choose those with $4 \leq \Delta r \leq 5$, i.e. factors of 40 to 100 in luminosity (and approximately in mass), a total of 597 objects (green in Fig. 3) out of the 11460 in the MASSES3-ISO sample. At low $z$ the magnitude range increases towards 10 mag. We therefore select an `extremely' isolated subsample (blue points in Fig. 3) with $5 \leq \Delta r \leq 10$ of 185 objects (which is only complete at the lower $z$ end, of course). M94 appears to be a local example of a galaxy with $\Delta r \sim 10$, or $10^{4}$ in mass ratio, see \cite{Smercina2018}. Galaxies closer to the boundary than these ($\Delta r$ below 4) may have significant (fairly high luminosity ratio) companions which are missed merely because of the proximity of the sample limit. It is therefore uncertain whether these are genuinely isolated or not (except in the sense that they do not have nearby galaxies brighter than or similar to themselves). 

\subsection{High luminosity ratio grouped galaxies}

We can also detour into the MASSES3-GROUP sample and search for any galaxies with large luminosity ratios relative to their companions, which is clearly possible for the most luminous primaries and faintest companions. For this we can use the GAMA group catalogue G3CFoFGroupv10 and use the listed $r$-band magnitude difference between the brightest galaxy and second brightest galaxy. Out of the 2982 first ranked galaxies in our groups sample, we find only 58 which are between 4 and 5 magnitudes brighter than any companions and only 11 more than 5 magnitudes brighter. With a brighter limiting magnitude, i.e. brighter than the actual detected companions, these would satisfy the conditions to be in the `isolated' samples. Unsurprisingly all are at redshifts less than $z=0.03$, where very faint companions are detectable, and virtually all are in galaxy pairs (putting more galaxies in the accessible magnitude range obviously pushes their magnitudes closer together). Despite the low numbers, the main point is that these galaxies are actually more isolated than many members of the MASSES3-ISO sample, that is, those less than 4 mag. above the sample limit. The lesson to take from this is, again, that apparent isolation depends on sample limits, hence our specific limits defined above and used in the next section.

\section{The Isolated Galaxies}

 From the previous section, it is clear that many of the MASSES3-ISO sample could have companions within a luminosity ratio of 1/40 which are undetected (in GAMA). We can therefore not guarantee that they are isolated in an absolute sense. However we do have galaxies which are definitely dominant in their group (if any); specifically we will explore the properties of the `very' ($4 \leq \Delta r \leq$ 5) and `extremely' isolated ($\Delta r \geq 5$) galaxies. We will refer to these subsets as ISO4 (597 galaxies) and ISO5 (185 galaxies). For comparison we create subsets of MASSES3-GROUP which we refer to as GROUP4 (1042 objects) and GROUP5 (493 objects) comprised of galaxies in the grouped sample which are the same numbers of magnitudes above the sample limit. Thus the only difference between ISO4 (or 5) and GROUP4 (or 5) galaxies is that the former have no companions in the 4 (or 5) mag. gap between themselves and the sample limit while the latter do. All other selection effects will be the same.

Fig. 4 shows a standard colour-magnitude diagram $M_{r,*}$ vs. $(u-r)_*$ \citep[e.g.,][]{Baldry2006}. The green and blue points are for ISO4 and ISO5 as in Fig. 3 and the grey points are for the combined GROUP4 and GROUP5 comparison samples.  It is evident that (both sets of) the isolated galaxies are preferentially in the blue cloud, $(u-r)_* \simeq 1$, while the grouped ones (at the same $z$ and apparent magnitude, by construction) have a large fraction of red galaxies at $(u-r)_* \simeq 2$. This is shown more clearly in Fig. 5; specifically, only 32\% (250/782) of the very or extremely isolated galaxies are redder than $(u-r)_* = 1.5$ (the centre of the GAMA low $z$ 'green valley'; see Bremer et al. 2018, Phillipps et al. 2019) compared to 55\% (841/1535) of the matched group galaxies. The variation of red galaxy fraction with environment is of course a very well known result \citep[][and many others; see Bhambhani et al. 2023 for a recent summary]{Kauffmann2004, Khim2015}. Here we have extended it specifically to the most isolated objects; recall that even the groups are mostly small, often pairs, GAMA containing no large clusters \citep{Robotham2011}. This systematic distinction even between isolated, pair and small group galaxies is seen in the current semi-analytic GAEA models \citep{Vulcani2026}.

On the other hand, the distributions of absolute magnitude are generally similar in form for ISO4 and GROUP4, peaking at $M_{r, *}$ around $-22.5$ in each case, as per Fig. 6. (The histograms here, as in what follows, are normalised by the numbers of objects in each sample). In both samples, the large majority of galaxies are brighter than $M_{r,*} \simeq -21$ (83\% for ISO4, 90\% for GROUP4), though there is a small 'excess' of brighter magnitudes for the grouped sample, the median being $-22.3$ compared to $-22.1$ for ISO4. The samples are statistically different at the 1\% level in a K-S test. (All distribution parameters are summarised in Table 1, below.) This shift is much greater between the ISO5 and GROUP5 subsets (Fig. 7) where the extremely isolated galaxies are significantly shifted to fainter magnitudes (median $M_{r,*} = -21.9$ compared to $-22.5$) relative to their comparison sample. (See also the blue points at the right in Fig. 4). This may contribute to, but given Fig. 6 for the numerically dominant ISO4 and GROUP4 subsets, is clearly not the only factor in very isolated galaxies being bluer. 

We should note that the relatively low luminosity very isolated galaxies reach down to $M_{r,*} \simeq -17$ (or around $10^{8} M_{\odot}$; see below), so at the upper end of what would be classed as dwarfs. Of course, even fainter galaxies in the overall MASSES3-ISO sample {\it may} be very isolated but are necessarily less than 4 mag. above the sample apparent magnitude limit (refer back to Fig. 1) so do not meet our criterion. Low luminosity and mass objects are expected to be particularly susceptible to environmental effects, but our samples are very small at these magnitudes.\footnote{Even though GAMA goes significantly deeper than many other samples, the requirement for our isolated galaxies to be well above the overall magnitude limit still leaves only a relatively small number of low luminosity galaxies in our samples.} Nevertheless, from Fig. 4 we can see that almost all of the very or extremely isolated lower-luminosity galaxies are blue (50 of the 54 fainter than $M_{r,*} = -20$), with the remainder at intermediate colours, ($u-r)_* \simeq 1.7$. Again this is consistent with the semi-analytical model of \cite{Vulcani2026}.

\begin{figure}
\includegraphics[width=\linewidth]{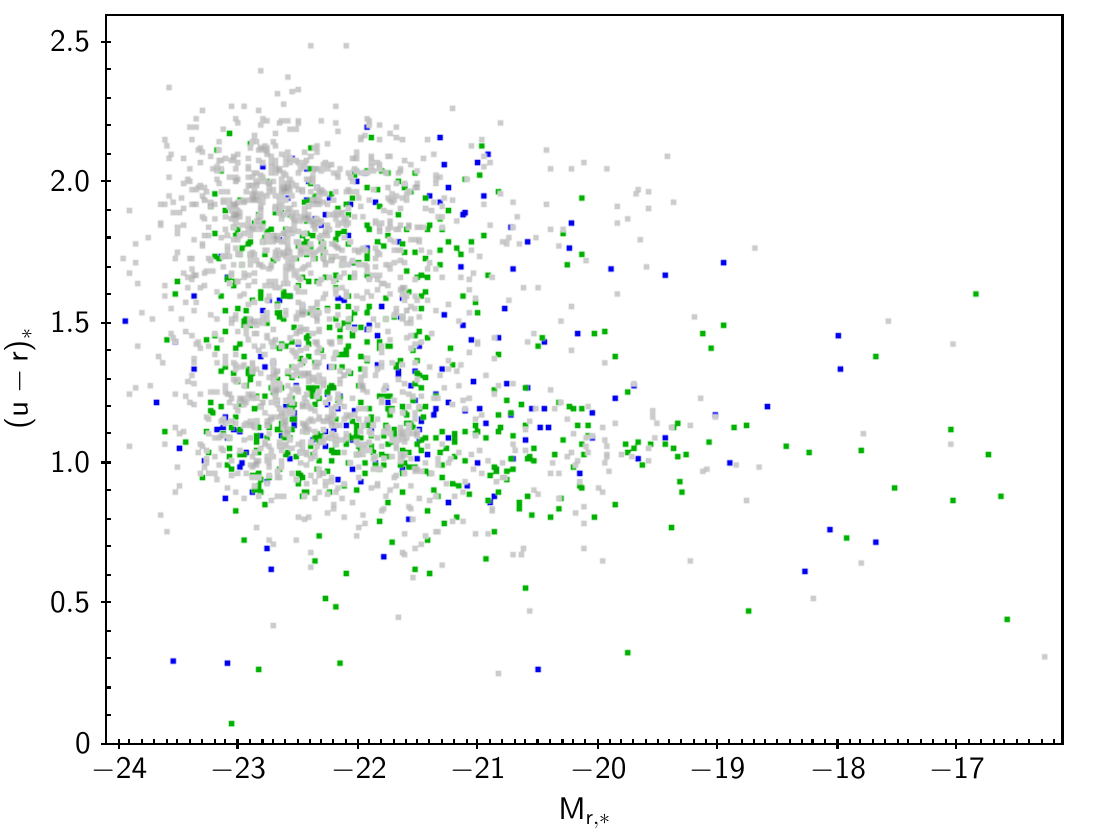}
\caption{Colour-magnitude plot for the subsamples of MASSES3-ISO. As before, galaxies with $4 < \Delta r < 5$ (sample ISO4) are shown in green and those with $\Delta r > 5$ (sample ISO5) in blue. The grey background points are for the (combined) comparison samples (GROUP4 + GROUP5) from MASSES3-GROUP, which have the same magnitude limits ($\Delta r > 4$).
}
\label{iso_col-mag}
\end{figure}

\begin{figure}
\includegraphics[width=\linewidth]{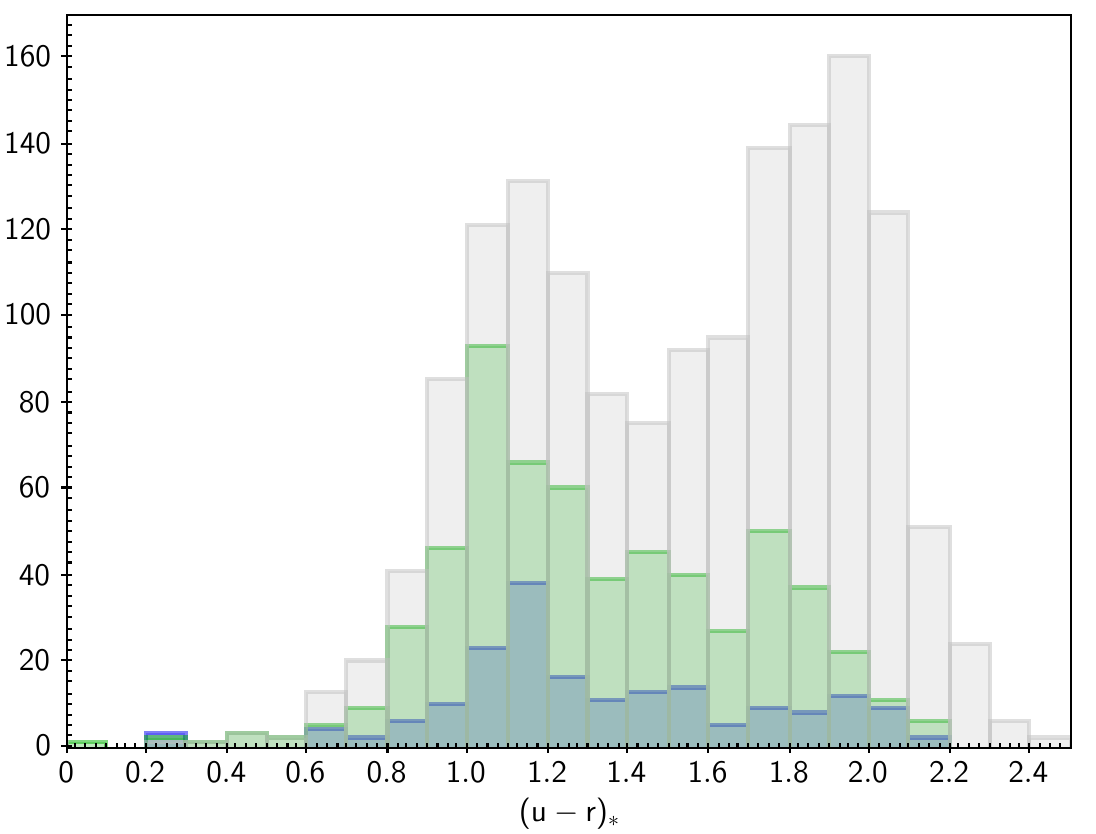}
\caption{The $(u-r)_*$ colour distribution for the samples in Fig. 4. ISO4 and ISO5 galaxies shown as the green and blue histograms, the comparison GROUP4 + GROUP5 galaxies in grey. 
}
\label{iso_col}
\end{figure}

\begin{figure}
\includegraphics[width=\linewidth]{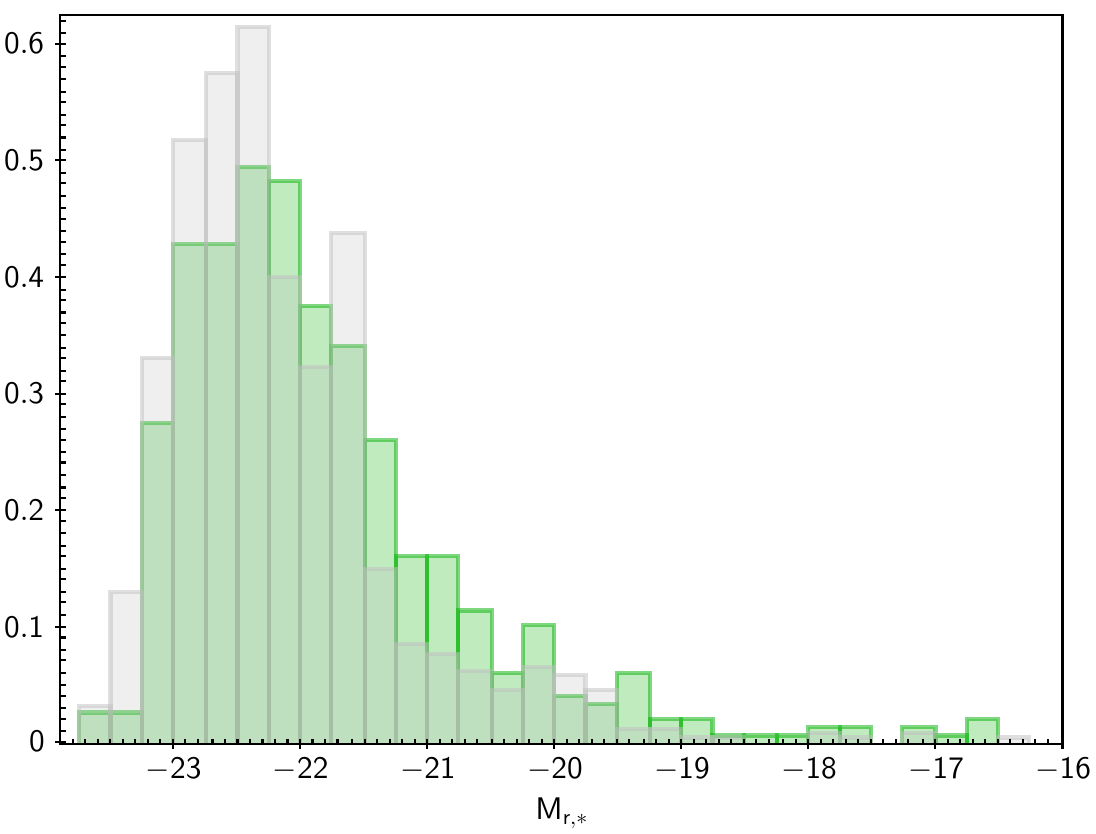}
\caption{The normalised absolute magnitude distribution for ISO4 (green) compared to GROUP4 (grey).
}
\label{iso4_mr}
\end{figure}

\begin{figure}
\includegraphics[width=\linewidth]{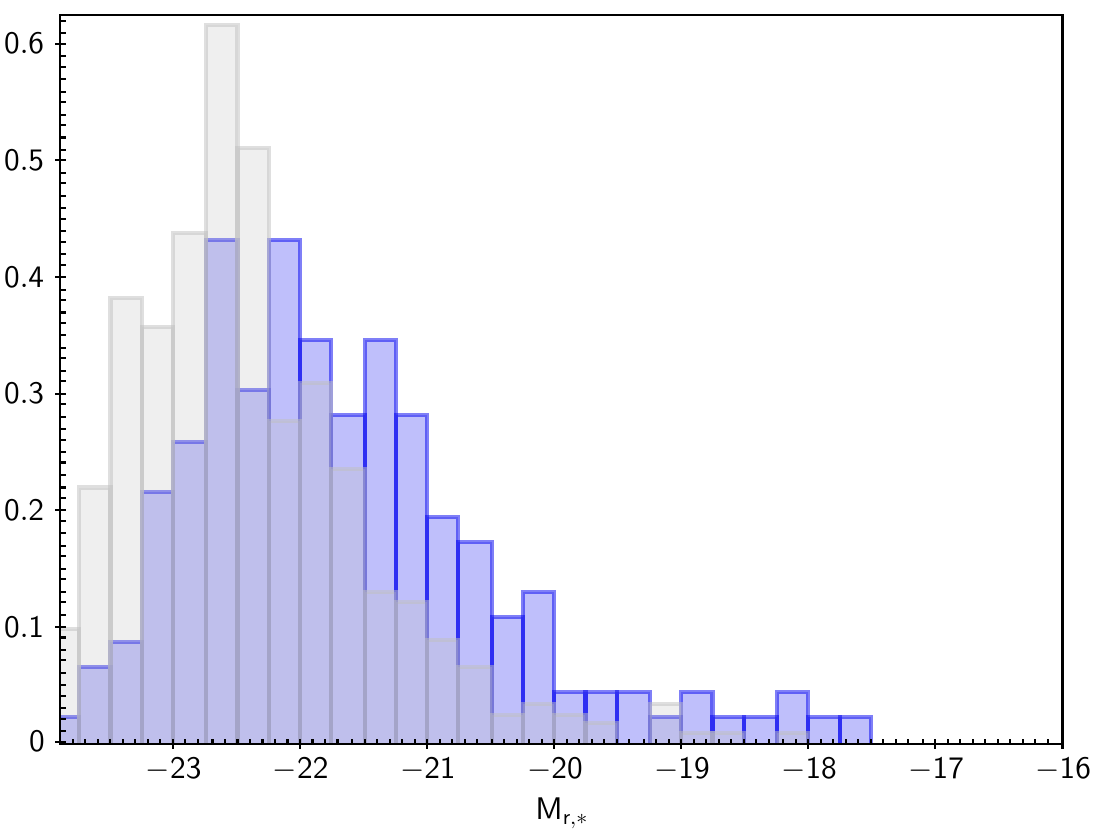}
\caption{As Fig. 6 but for ISO5 (blue) compared to GROUP5 (grey). 
}
\label{iso5_mr}
\end{figure}

Notice particularly that our result is a stronger statement than saying that the full `isolated' sample preferentially contains bluer galaxies than the `grouped' sample, since the former contains many galaxies for which significantly sized companions could not be detected in our data. Going to fainter apparent (and therefore absolute) magnitudes might move them from the ungrouped to the grouped samples. On the other hand, any unseen companions of our ISO4 (5) galaxies will necessarily be more than 40 (100) times less luminous than our sample galaxies, while GROUP4 (5) galaxies do have companions at least this bright: extending the magnitude limit would not change our classifications as defined in this way.

Next, Fig. 8 shows the Sersic indices of the different samples, taking the single Sersic fits from the GAMA DMU SersicPhotometryv09 \citep{Kelvin2012}. Consistent with the above, the `very' and `extremely' isolated galaxies largely have $r$-band Sersic indices around $n_r = 1$ to 2, implying (near) exponential discs, while the grouped galaxies (of the same magnitudes) have a second peak around $n_r =4$, indicating elliptical-like structures. Splitting the sample at $n_r = 3$, for instance, only one third of ISO4 and 5 galaxies (257/782) have spheroid-like $n_r$ compared to half of GROUP4 and 5 galaxies (770/1535). The plot of the central surface brightness of the fitted profiles (not shown) has the equivalent bimodal form for the grouped galaxies (both high and low central intensity) and unimodal but extended (mainly lower central intensity) for the isolated galaxies. Thus we have the very clear implication that the most isolated galaxies are morphologically primarily late types. In addition, as we would then expect, the MAGPHYS SED fitting shows that, overall, ISO4 and 5 galaxies are slightly less metal rich (and marginally dustier) than the corresponding GROUP galaxies.  

Given this, it is not surprising that the ISO4 and 5 galaxies are less massive (or rather, their mass distribution is shifted towards lower masses) than the GROUP4 and 5 galaxies, as seen in Fig. 9. (A K-S test gives less than 0.1\% probability that the two distributions are the same). This is shown in a different way in Fig. 10, where the MAGPHYS-derived mass is plotted as a function of $n_r$. Here, 77\% of galaxies with $M_* \geq 10^{10} M_{\odot}$ and $n_r \geq 3$ (top right of the plot) are in groups, compared to 50\% for galaxies below those limits (bottom left). In other words, the denser environments contain more high-mass, elliptical-like galaxies, as we would expect. Notice too, again as expected, that the lowest mass galaxies in both our samples (less than $10^9 M_{\odot}$) are virtually all at low $n_r$. 

\begin{figure}
\includegraphics[width=\linewidth]{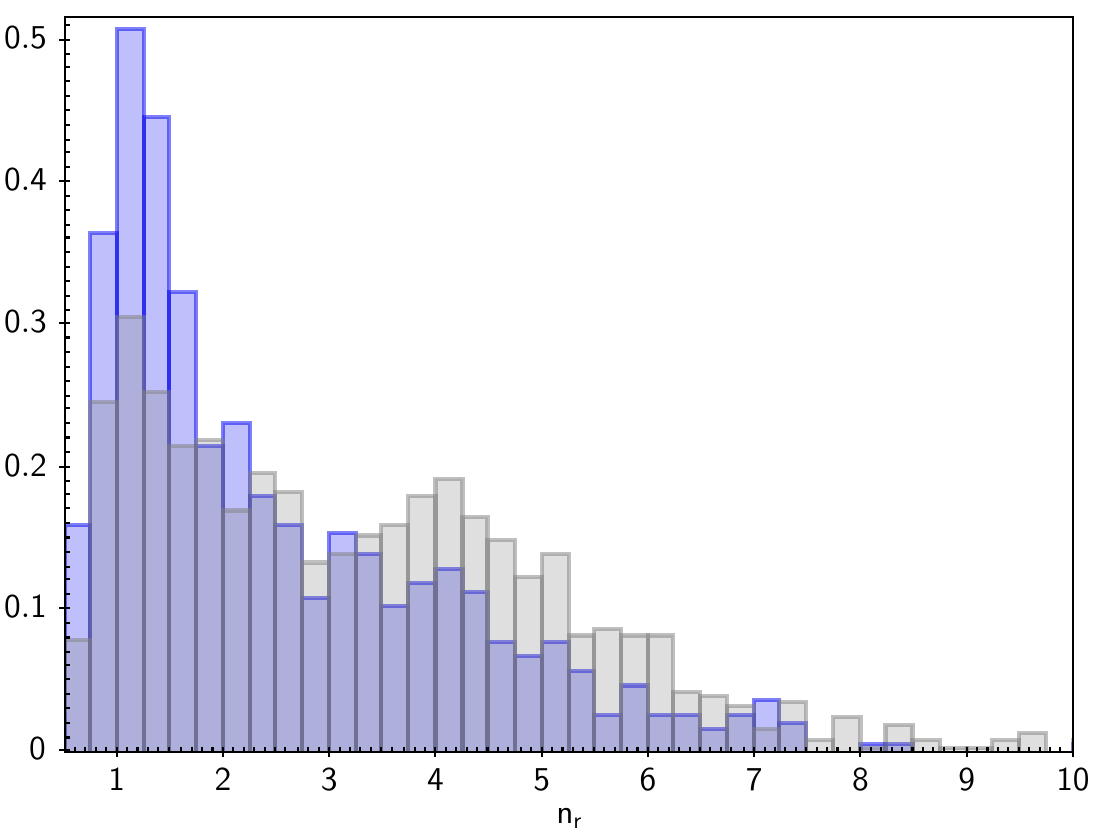}
\caption{The normalised $r-$band Sersic index distribution for the samples ISO4+ISO5 in blue, the comparison GROUP4+GROUP5 galaxies in grey. 
}
\label{iso_sersicn}
\end{figure}

\begin{figure}
\includegraphics[width=\linewidth]{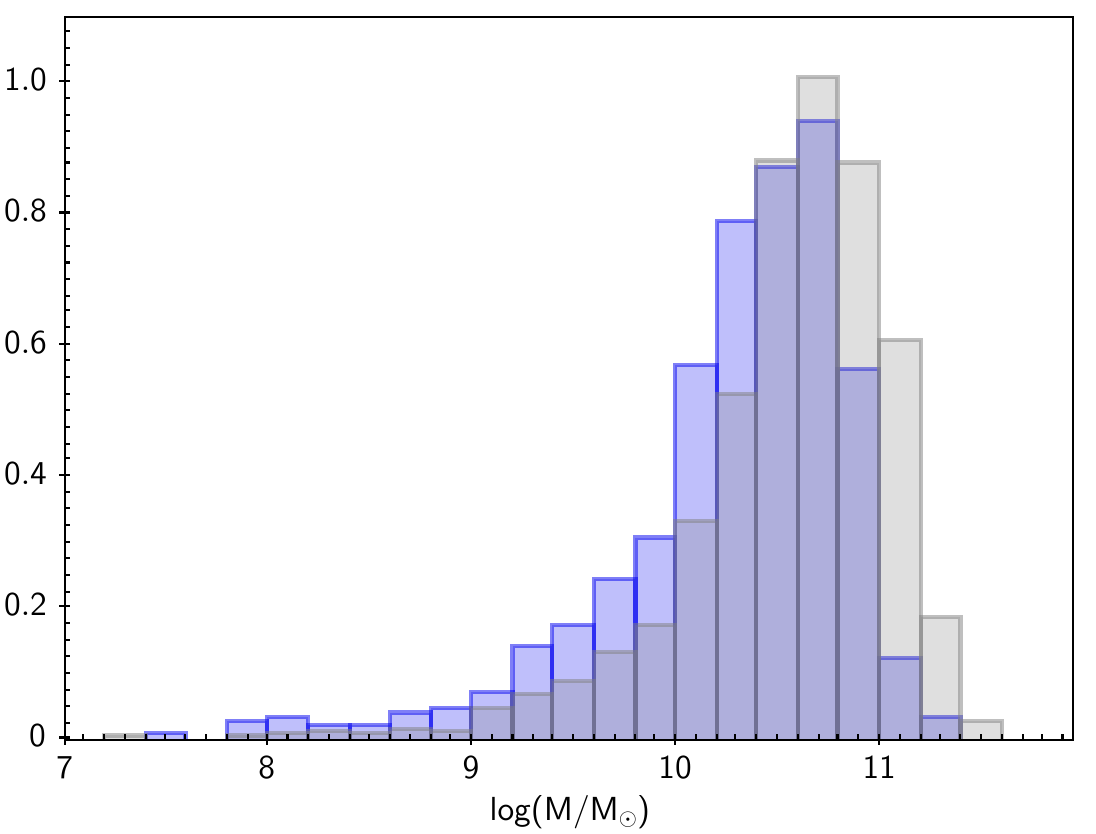}
\caption{As Fig. 8 but for stellar mass in units of $M_{\odot}$. ISO4+ISO5 galaxies in blue, GROUP4+GROUP5 galaxies in grey. }
\label{iso_mass}
\end{figure}

\begin{figure}
\includegraphics[width=\linewidth]{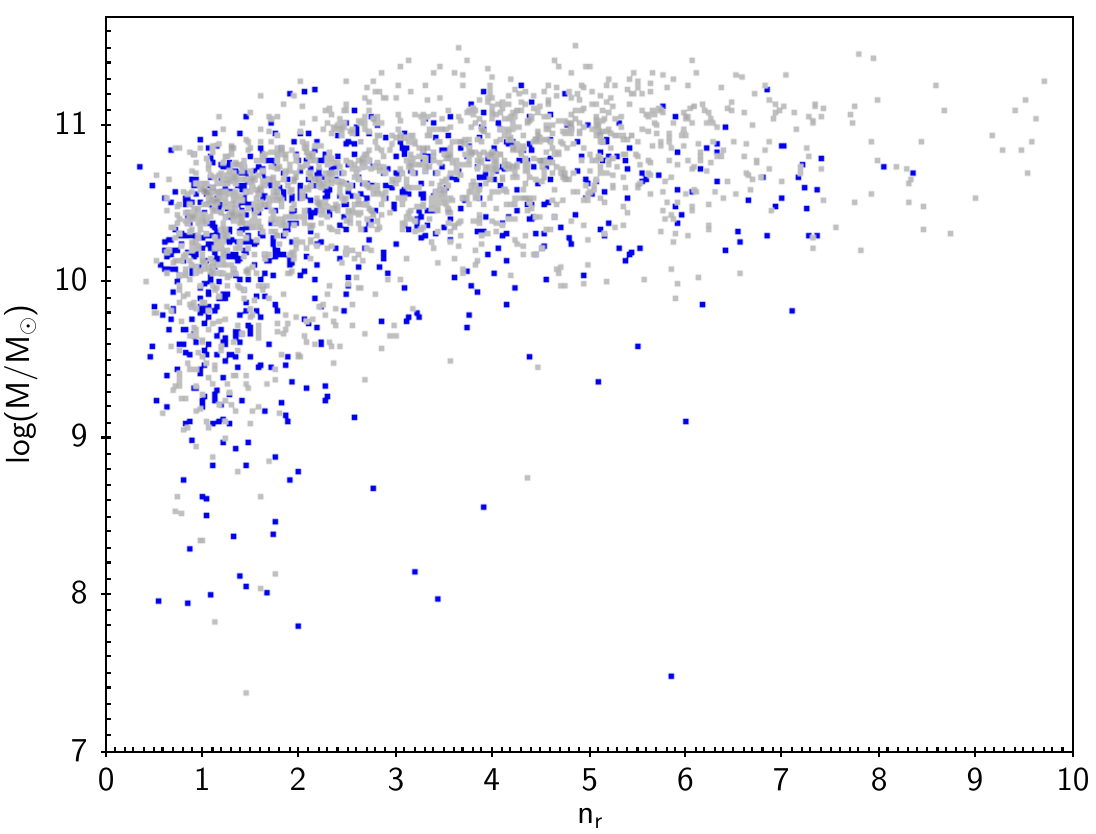}
\caption{The combined distribution of ISO4 and ISO5 galaxies (blue points) compared to GROUP4 and GROUP5 galaxies (grey points) in the mass-Sersic index plane.}
\label{iso_mass_nr}
\end{figure}

Moving on to the stellar population parameters, Fig. 11 shows a similar effect in the $r$-band light-weighted mean stellar age from the MAGPHYS analysis, the ISO4 and ISO5 galaxies have a spread of ages around $10^{9.4}$Gyr while the corresponding GROUP samples peak at older ages around $10^{9.8}$Gyr \citep[cf.][]{Torres2024}. The difference is even clearer in the fitted e-folding time of the star formation rate, $\tau$, in Fig. 12. Both ISO and GROUP samples show bimodal distributions but the ISO4 and 5 galaxies are weighted much more strongly to the high $\tau$, i.e. long lasting star formation, than the GROUP4 and 5 galaxies, which populate the two peaks, for long lasting and short burst of star formation, approximately equally. For instance, splitting the data at log($\tau$/yr) = 8.8, 55\% of grouped galaxies have short timescales compared to only 32\% of isolated galaxies.

\begin{figure}
\includegraphics[width=\linewidth]{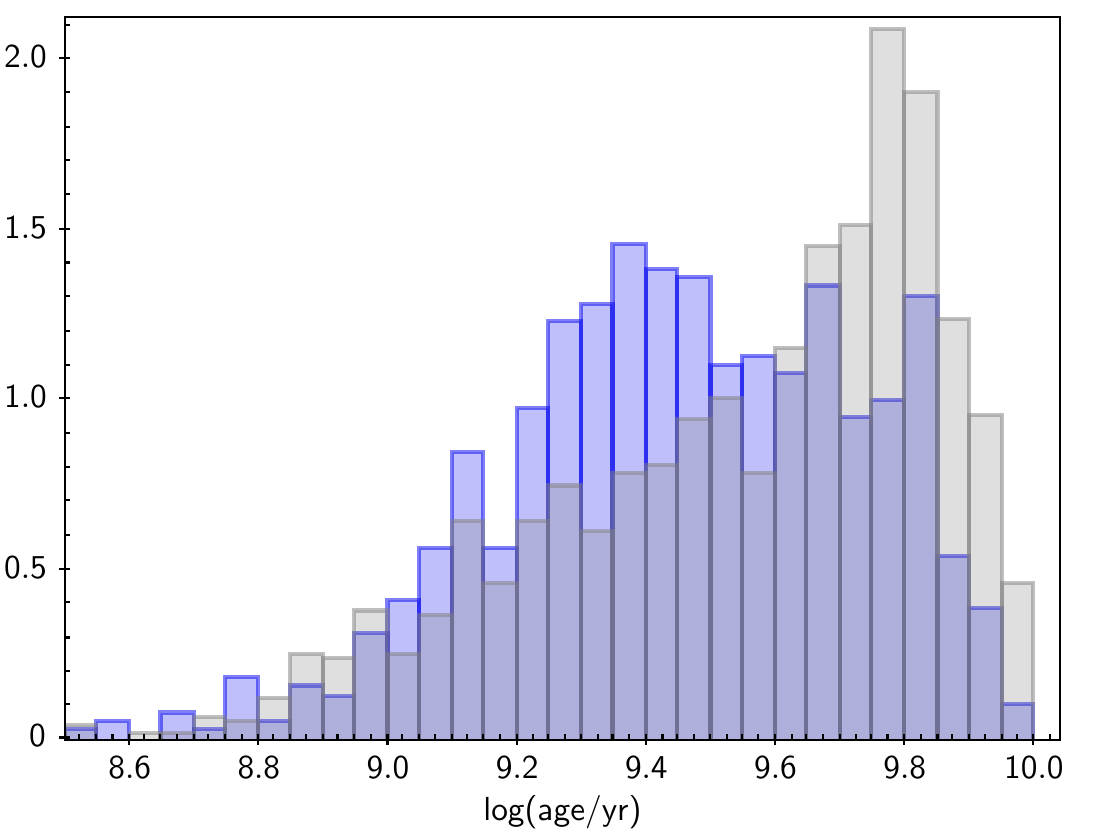}
\caption{As Fig. 8 but for the light-weighted age of the stellar population. ISO4+ISO5 galaxies in blue, GROUP4+GROUP5 galaxies in grey. 
}
\label{iso_lwage}
\end{figure}

\begin{figure}
\includegraphics[width=\linewidth]{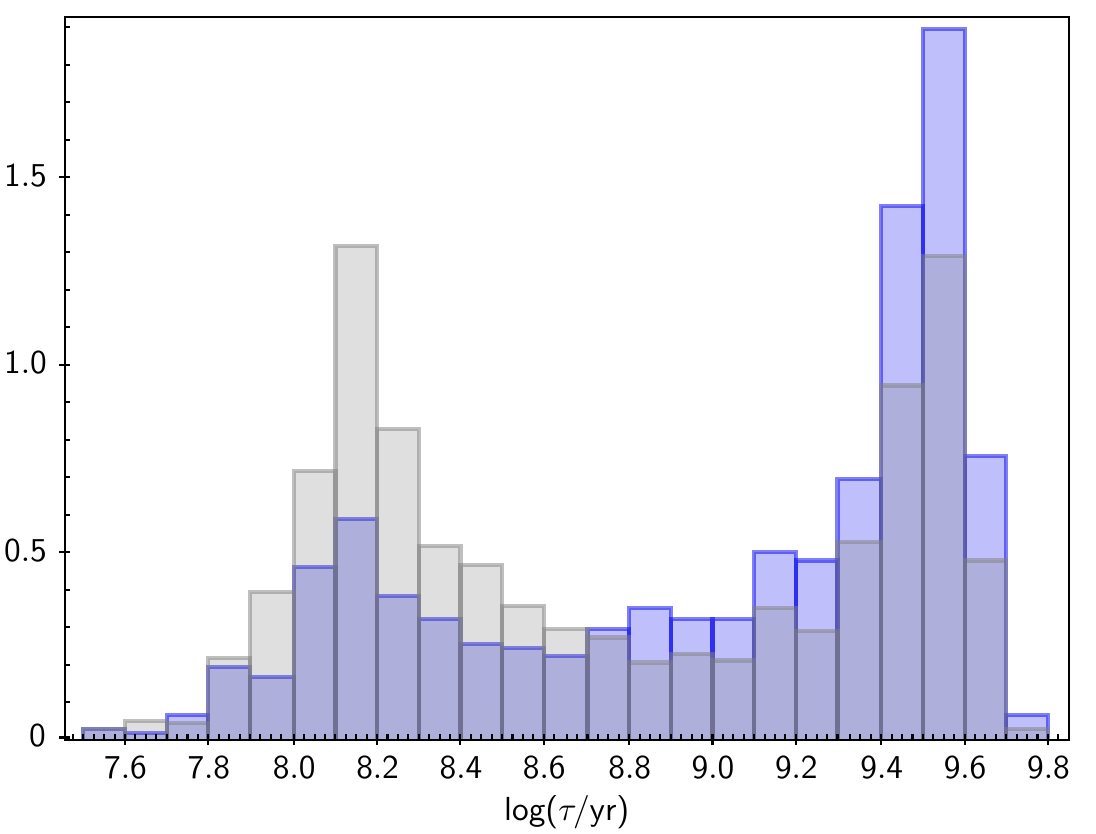}
\caption{As Fig. 8 but for the star formation e-folding time, $\tau$. ISO4+ISO5 galaxies in blue, GROUP4+GROUP5 galaxies in grey.
}
\label{iso_tau}
\end{figure}

In terms of the current (or recent) star formation rate itself, Fig. 13 shows the specific SFR (SFR per unit stellar mass) averaged over the last $10^7$ years. As would be expected from the foregoing, the GROUP4 and 5 galaxies are essentially equally divided between low ($\simeq 10^{-12}$yr$^{-1}$) and moderately high ($\simeq 10^{-10.5}$yr$^{-1}$) sSFR (50\% are above $10^{-11}$yr$^{-1}$), whereas the ISO4 and 5 galaxies are primarily (73\%) in the higher sSFR peak \citep[cf.][]{Davies2016b}. Averages over other time scales give the same result.

\begin{figure}
\includegraphics[width=\linewidth]{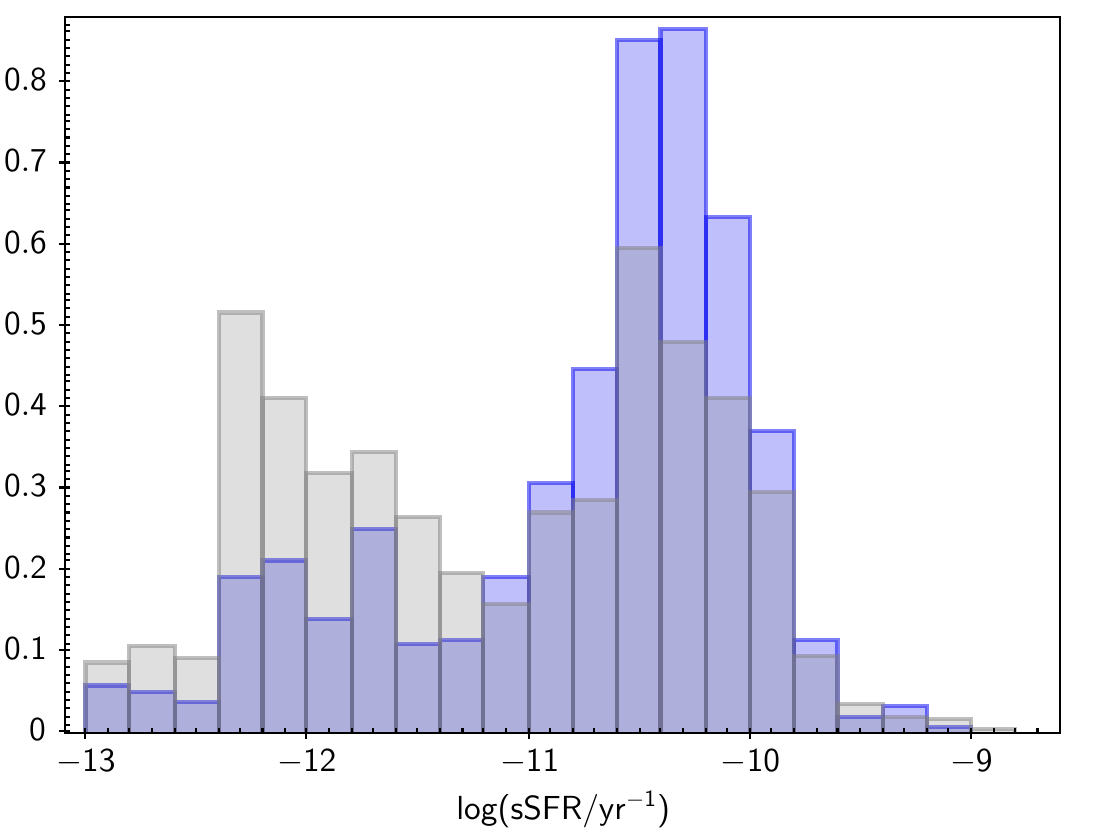}
\caption{As Fig. 8 but for the specific star formation rate averaged over the last $10^{7}$ years. ISO4+ISO5 galaxies in blue, GROUP4+GROUP5 galaxies in grey.
}
\label{iso_sfr}
\end{figure}

\section{Discussion}

The clear result of the previous section is that galaxies which are definitely isolated from any others within a factor of 40 of their luminosity are more likely to be blue, disk-like galaxies with extended and ongoing star formation, and hence lower mean stellar ages, than are galaxies of the same $r$-band magnitude at the same redshift which do have companions with at least 1/40 of their luminosity. Note, though, that these are not clean divisions. Unsurprisingly, the grouped galaxies span both red, passive, spheroidal and blue, star-forming, disk galaxies, but so do the isolated ones, though with clearly different relative numbers.

Thus, in terms of global evolution, it appears that isolation {\it favours}, but does not enforce, the processes which give rise to (or maintain) disk-like morphologies and prolong star formation. The latter cannot currently be (re)fuelled by infall of material from companions or be promoted by interactions, though, of course, either of these may have happened in the past via now dissolved satellites. 

On the other hand, it has been possible for {\it some} galaxies to evolve to early-type forms in apparent isolation, with the same caveat about earlier events.\footnote{The red very isolated galaxies could, for instance, be related to the massive galaxies which have apparently cannibalised their neighbours to form `fossil groups' \citep{Ponman1994, Khosroshahi2014}. However, there is no evidence for possible signatures of this process such as particularly high masses or high dust content in our sample of isolated red/early-type galaxies}. \cite{Hirschmann2013} suggested from theoretical models that the large majority of isolated galaxies should have been isolated since at least $z=1$, but that a small percentage will be bulge-dominated systems built up by previous mergers. Interestingly, referring back to Fig. 4, there appear to be no very isolated red, low-luminosity galaxies. This is consistent with the idea that low-mass galaxies remain star-forming unless affected by their environment \citep[][and references therein]{Geha2012, Davies2016b}.

In any case, given that the fraction of red/quiescent galaxies is clearly different between our matched isolated and grouped galaxies, even though most of the latter are in very small groups, the production of such objects was presumably easier in these denser environments \citep[cf.][]{Melnyk2015}. Specifically, it can be argued that the shortage of red galaxies in the very isolated sample points to close interactions as a key factor in their production. 

Further, the relatively few galaxies in Figs. 5 and 13 seen transiting through intermediate colour or SFR (the green valley) suggests that red galaxy formation was generally rapid and/or early, as discussed in \cite{Brough2013} and \cite{Kinyumu2024}, amongst others. In general, as would be expected, in both our environments red galaxies have the shortest SFR e-folding times and green galaxies slightly longer ones with $\tau \sim 10^{8.8}$yr, as reflected in Fig. 12 \cite[see][for a more detailed discussion]{Phillipps2019}. Thus the large difference in red fraction between our grouped and very isolated galaxies can be taken to imply that some processes related to interactions expedited the early cessation of star formation even in the relatively low density environment of small groups. Recall that, on average, grouped galaxies had shorter star-formation e-folding times and older stellar populations (consistent with the larger fraction of early type galaxies.

Moving on specifically to currently star-forming galaxies, we can compare the very isolated galaxies to grouped galaxies of the same mass. Previous works have reached differing conclusions as to their star formation. For instance, \cite{Argudo2025} suggest that for their SDSS samples, isolated galaxies have, on average, slightly {\em lower} sSFR and a tighter `main sequence of star-forming galaxies' (MSSF) in the mass-sSFR plane \citep{Noeske2007, Speagle2014} than do those in pairs and triplets. Conversely, \cite{Erfanianfar2016} previously found isolated spirals to have {\em higher} sSFR than other galaxies, while Melnyk et al. (2015) suggested that any difference between isolated and paired galaxies was mass dependent. Using GAMA data, \cite{Grootes2017} found group galaxies to be statistically indistinguishable from `field' galaxies in sSFR. (See also \cite{Spector2017} who studied a small sample of `extremely isolated galaxies'). We should be aware, here, that selection criteria vary significantly between different authors. In the particular cases quoted above, for example, Argudo-Fernandez et al., Grootes et al. and Erfanianfar et al. utilise progessively deeper samples and higher redshift ranges; Argudo-Fernandez et al. also select star-forming galaxies via emission lines while the others use multi-wavelength photometry. In addition, of course, as discussed earlier, definitions of `isolated' frequently differ between authors.

To investigate the star formation distributions for our samples in more detail, we select objects with masses between $10^{8.5}$ and $10^{11} M_{\odot}$ and sSFR above $10^{-11}$yr$^{-1}$ (i.e., to the right of the green valley minimum in Fig. 13) as being on, or close to, the main sequence of star formation. The isolated and grouped `star-forming' galaxies in this sample show almost identical sSFR distributions in Fig. 14. (The means and standard deviations are $-10.34 \pm 0.31$ and $-10.32 \pm 0.34$, respectively, and a K-S test gives no evidence of a difference, even at the 20\% level). More specifically, they have the same distribution in the mass-sSFR plane (Fig. 15); from the density contours we can see that both samples fall along the same main sequence and with the same scatter. Although the log(sSFR)-log(mass) relation is non-linear \citep[][and references therein]{Davies2025,He2026}, as a reasonable, quantifiable representation of the visible trend of the MSSF across our samples' ranges, we can note that the best fit slopes in Fig. 15 are $-0.30$ for the ISO galaxies and $-0.34$ for the GROUP galaxies. The effective standard deviations (enclosing 68\% of the points), around the respective trend lines, are $\pm 0.27$ and $\pm 0.28$.

As the grouped sample can contain both first ranked (brightest group member) and satellite galaxies, which may differ in some aspects of the star formation process, it is worth considering whether this affects the comparison with isolated galaxies. In our GROUP4 and 5 sample, two thirds of the galaxies are `centrals' and one third are satellites. Their separate sSFR distributions are very similar, and both are very similar to the isolated galaxies above, with mean and standard deviation $-10.30 \pm 0.33$ for centrals and $-10.36 \pm 0.35$ for the satellites. (Recall that all these star-forming samples are limited at sSFR = $10^{-11}$yr$^{-1}$).

\begin{figure}
\includegraphics[width=\linewidth]{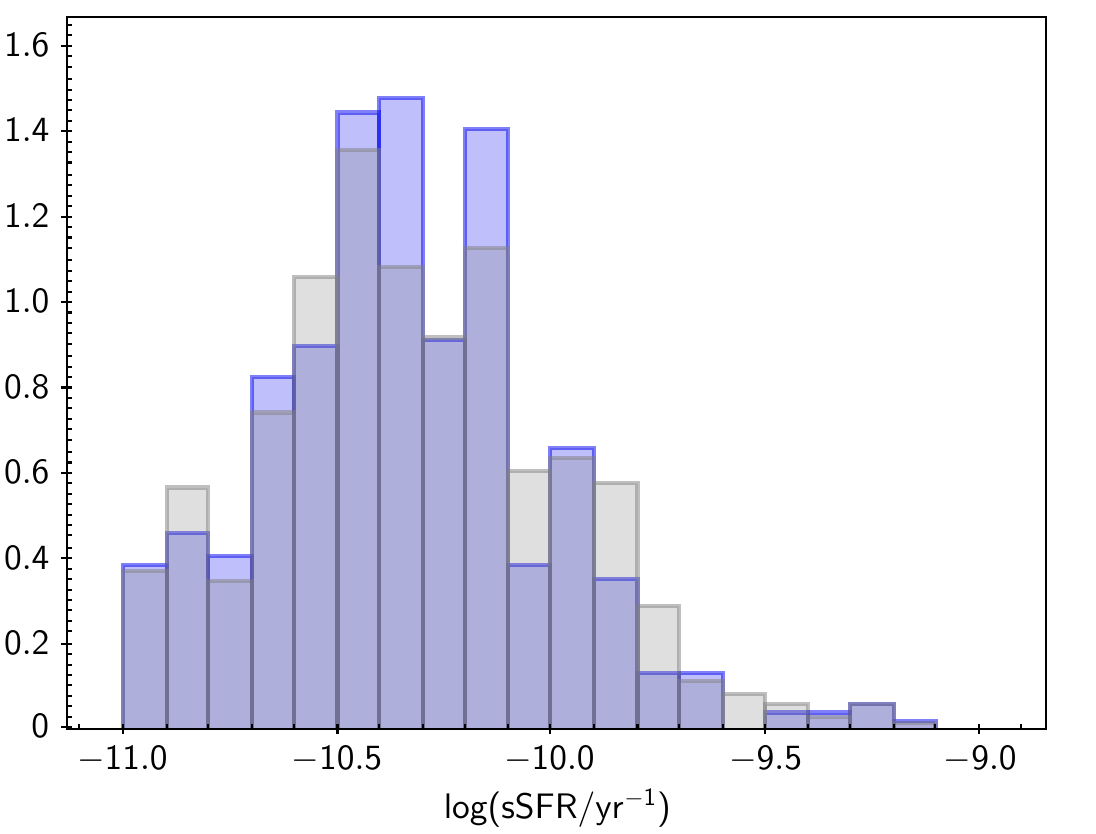}
\caption{The renormalised distribution of sSFR for galaxies with sSFR above $10^{-11}$yr$^{-1}$ ('star-forming galaxies') only. Star-forming ISO4+ISO5 galaxies in blue, star-forming GROUP4+GROUP5 galaxies in grey.
}
\label{iso_ssfr11}
\end{figure}

\begin{figure}
\includegraphics[width=\linewidth]{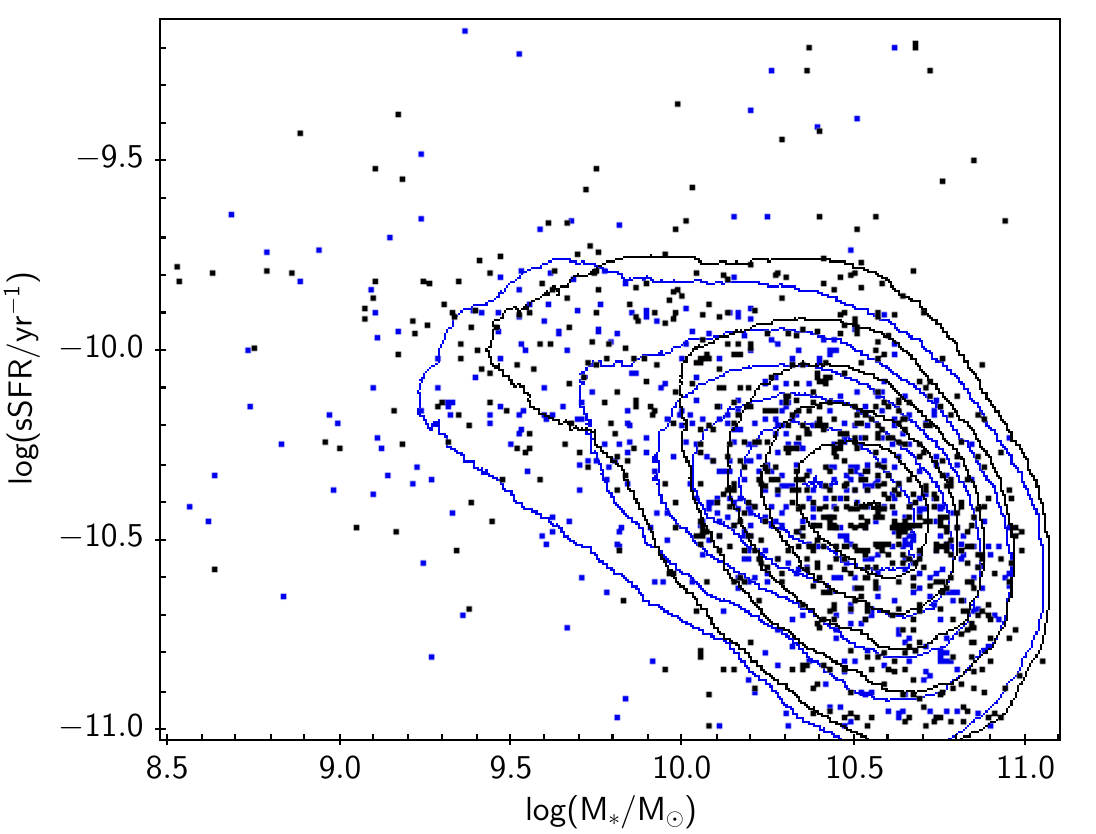}
\caption{The distribution of star-forming ISO4+ISO5 galaxies (blue points), compared to star-forming GROUP4+GROUP5 galaxies (grey points) in the sSFR-mass plane, with contours of number density superimposed.
}
\label{iso_ssfr11_mass}
\end{figure}

Equivalent plots of mass against $(u-r)_*$ or against light-weighted age are again the same for each sample. Recall that for the overall ISO4 and 5 and GROUP 4 and 5 samples, the latter showed significantly older ages (Fig. 11), but once the samples are restricted to only currently star-forming galaxies, the two histograms become very similar with a strong peak at $10^{9.2}$ to $10^{9.6}$~yr (Fig. 16). Though the distribution for the grouped galaxies is slightly broader (log(age) $= 9.31 \pm 0.28$, compared to $9.35 \pm 0.22$), a K-S test shows no significant difference in the overall distributions at the 10\% level). Similarly, the distributions of e-folding time $\tau$ are now very similar for the star-forming ISO4 and 5 and GROUP4 and 5 galaxies, with a sharp peak at $10^{9.5}$~yr.

\begin{figure}
\includegraphics[width=\linewidth]
{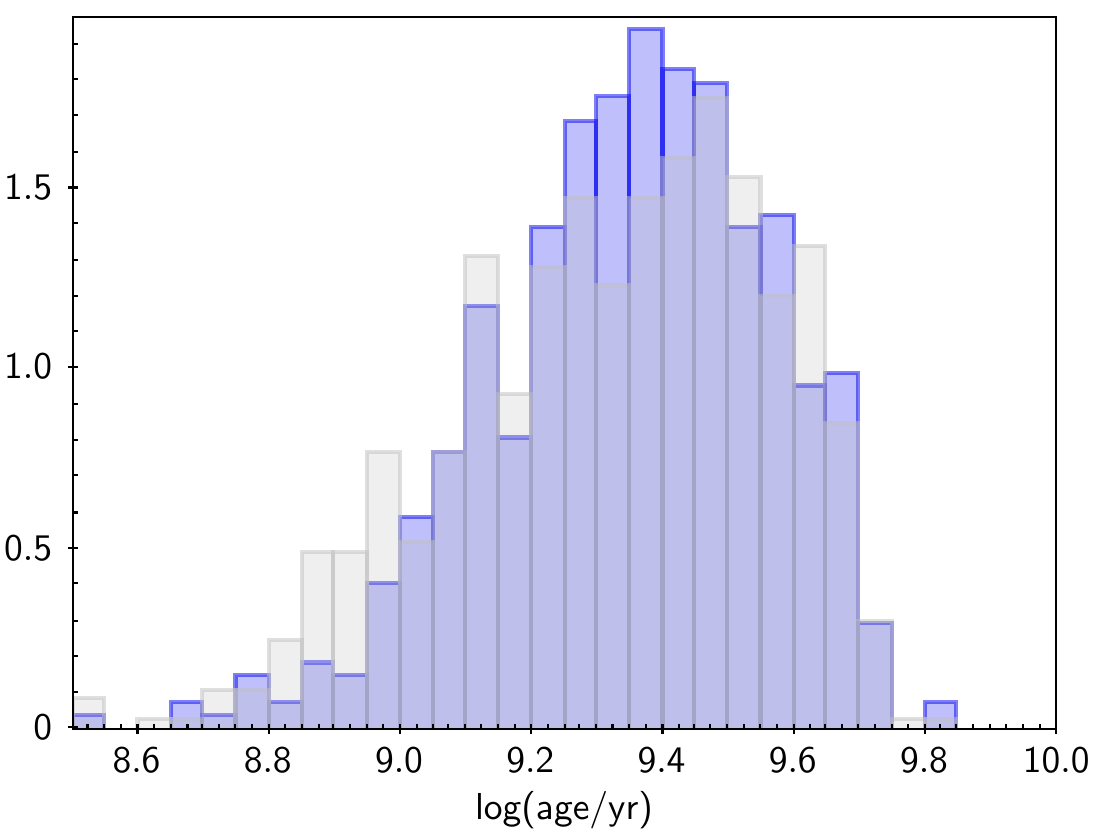}
\caption{As Fig. 14 but for luminosity-weighted age. ISO4+ISO5 star-forming galaxies in blue, GROUP4+GROUP5 star-forming galaxies in grey.
}
\end{figure}

The dust-to-stellar mass ratio $M_d/M_*$ \citep[which can be treated as a reasonable proxy for gas fraction;][]{Eales2010, Phillipps2023}, also has the same distribution for the star-forming isolated and grouped galaxies (Fig. 17). \cite{Roychowdhury2022} previously found the equivalent result for directly measured HI fractions of galaxy pairs or groups versus isolated galaxies of the same mass, using high resolution DINGO observations in one of the GAMA fields. \citep[There is no evidence in the present data, or in Roychowdhury et al., for the narrower distribution for isolated compared to grouped galaxies suggested by][]{Bok2020}. 

\begin{figure}
\includegraphics[width=\linewidth]{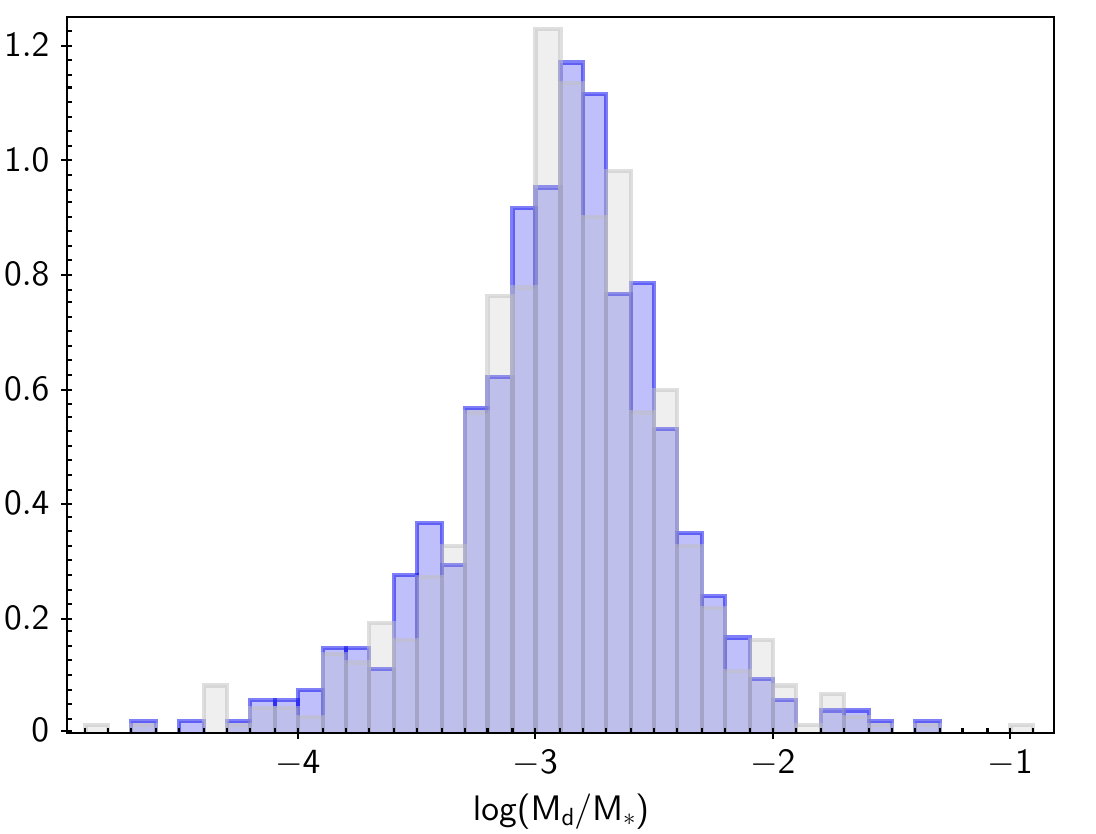}
\caption{As Fig. 14 but for the dust-to-stellar mass ratio. ISO4+ISO5 star-forming galaxies in blue, GROUP4+GROUP5 star-forming galaxies in grey.
}
\label{iso_dust}
\end{figure}

\begin{table}
\begin{tabular}{| c c c c c |}
\hline
  & ISO4 & GROUP4 & ISO5 & GROUP5 \\ 
 Median $M_r$ & $-22.1$ & $-22.3$ & $-21.9$ & $-22.5$ \\ 
 \hline
  Star Forming& ISO4+5 & GROUP4+5& Central & Satellite \\  
 log(sSFR) & $-10.34$ & $-10.32$ & $-10.30$ & $-10.36$ \\ 
  & $\pm$ 0.31 & $\pm$ 0.34 & $\pm$ 0.33 & $\pm$ 0.35\\
  log(age) & 9.35 & 9.31 & & \\
  & $\pm$ 0.22 & $\pm$ 0.28 & & \\
 MSSF slope & 0.30 & 0.34 & & \\
 scatter &0.27 & 0.28 & & \\
 \hline
\end{tabular}
\caption{Summary of distribution parameters for isolated and grouped subsets. Lower section for star-forming galaxies only.}
\label{table:1}
\end{table}

The scatter plots of sSFR versus $M_d/M_*$ (effectively a rough measure of star formation efficiency, as a scaled version of SFR versus gas mass) are also the same for the two samples, as shown by the contours in Fig. 18. 

\begin{figure}
\includegraphics[width=\linewidth]{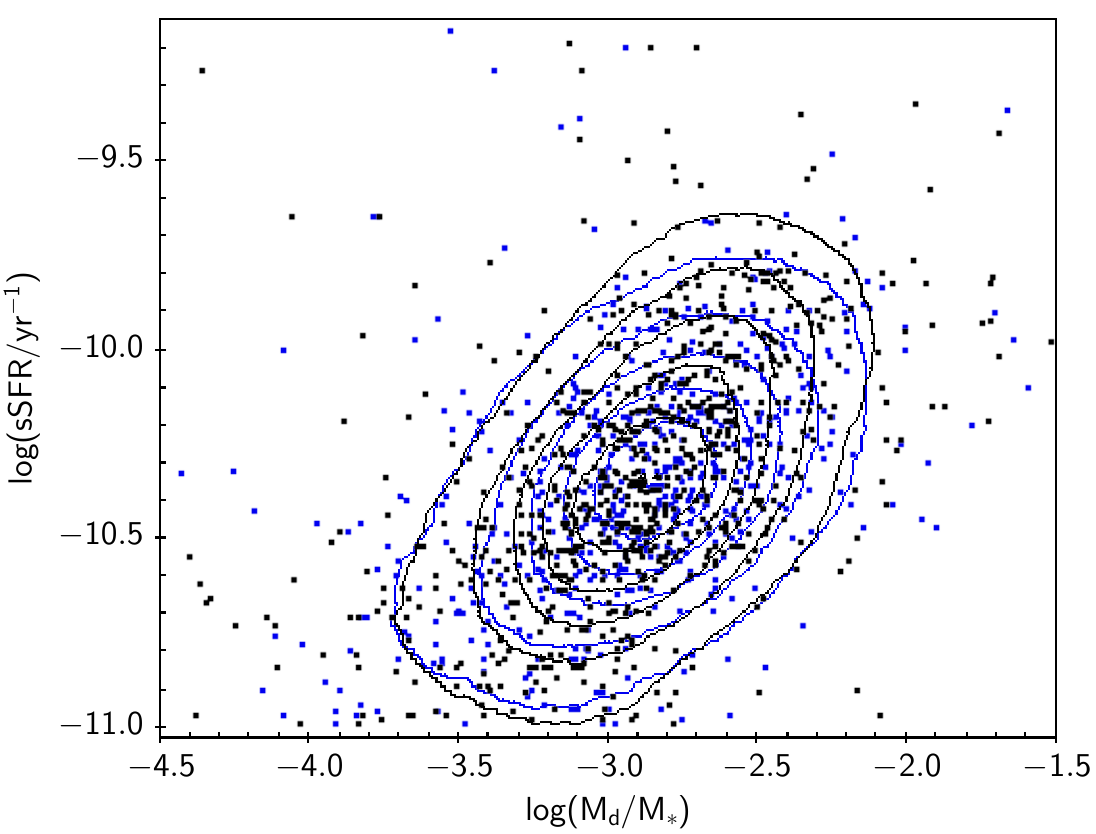}
\caption{As Fig. 15 but for the sSFR-dust mass plane. ISO4+ISO5 star-forming galaxies in blue, GROUP4+GROUP5 star-forming galaxies in grey.
}
\label{iso_sfr_dust}
\end{figure}

We therefore conclude that, in our well defined, carefully matched samples, there appear to be no significant differences in the recent star formation processes in the galaxies with significant companions and those without. This extends, to the lowest possible density regime, the conclusions of previous work \citep[cf.][]{Balogh2004, Peng2010, Wijesinghe2012, Brough2013}, which found the same star formation characteristics for galaxies in different environments. \citep[See also the recent theoretical work on galaxies in voids by][]{Schiller2026}. Given the lack of significant interactions for the isolated sample, this implies that the evolution of present day star-forming galaxies in all environments (or at least those outside rich clusters) has been, and remains, dominated by internal processes.

\section{Summary and Conclusions}

We have used GAMA survey data and an objective measure of isolation for nearby ($z \leq 0.1$) non-group galaxies, in terms of a minimum separation in magnitude between the galaxy and the survey limit (or equivalently the maximum possible luminosity ratio of any unseen companion), to find the most isolated galaxies in the survey. We have then compared the physical characteristics of our very isolated galaxies to those of comparison samples of grouped galaxies with the same magnitude limits.

We find that, as expected, while grouped galaxies split approximately equally between between blue, disk-like, currently star forming galaxies and red, spheroidal, quiescent galaxies, isolated galaxies are preferentially (two thirds or more) the former. Virtually all low-luminosity isolated galaxies are blue.  The overall isolated sample also contains similarly larger fractions of galaxies with long star formation time scales and young stellar population ages. These differences occur despite the fact that more than half of our grouped objects are merely in pairs and triples, rather than anything larger.

Notwithstanding the decrease in red fraction, it is evident that some (currently) isolated galaxies have managed to become spheroidal and quiescent, possibly due to effects occurring at early times.

Considering only the star-forming galaxies, when controlled by mass we find no differences between the grouped and isolated galaxies in terms of their star formation, that is, they exibit closely similar distributions along the main sequence of star formation. Isolated and grouped star-forming galaxies also possess very similar distributions of star formation time-scale and mean stellar age. Further, the dust-to-stars mass ratio, a proxy for gas fraction, tracks specific star formation in the same way for both isolated and grouped galaxies, suggesting that the star formation efficiency is the same in each case, i.e. independent of the presence of companions.

Given the lack of significant interactions for the isolated sample, these results appear to imply that the evolution of present-day star-forming galaxies in all environments (at least outside rich clusters) has been, and remains, dominated by internal (secular) processes. That grouped star-forming galaxies, which can obviously have been affected by interactions, look exactly like those with no companions may suggest that interactions in groups are an `all or nothing' occurrence. The galaxy concerned may either be `quenched' and pushed towards the red sequence (hence the larger red fraction) or, perhaps in the case of smaller perturbations, quickly recover and resume its place on the main sequence.

\section*{Acknowledgements}

GAMA is a joint European-Australasian project based around a
spectroscopic campaign using the Anglo-Australian Telescope.
The GAMA input catalogue is based on data taken from the Sloan Digital Sky Survey and the UKIRT Infrared Deep Sky Survey. Complementary imaging of the GAMA regions is being obtained by a number of independent survey programmes including GALEX MIS, VST KiDS, VISTA VIKING, WISE, Herschel-ATLAS, GMRT and ASKAP providing UV to radio coverage. GAMA was funded by the
STFC (UK), the ARC (Australia), the AAO, and the participating institutions. The GAMA website is http://www.
gama-survey.org/ . This paper has made extensive use of the TOPCAT software package \citep{MBTaylor}.

\section*{Data Availability}
The data underlying this paper are available at http://www.gama-survey.org/dr4/.




\end{document}